\documentclass[twocolumn, switch]{article} 

\usepackage{amsmath, amsthm, amssymb, amsfonts}

\usepackage[numbers,square]{natbib}
\usepackage[utf8]{inputenc}	
\usepackage[T1]{fontenc}	
\usepackage{xcolor}		
\usepackage[colorlinks = true,
            linkcolor = purple,
            urlcolor  = blue,
            citecolor = cyan,
            anchorcolor = black]{hyperref}	
\usepackage{booktabs} 		
\usepackage{nicefrac}		
\usepackage{microtype}		
\usepackage{lineno}		
\usepackage{float}			

\usepackage{url}

\usepackage{newfloat}
\DeclareFloatingEnvironment[name={Supplementary Figure}]{suppfigure}
\usepackage{sidecap}
\sidecaptionvpos{figure}{c}

\usepackage{titlesec}
\titlespacing\section{0pt}{12pt plus 3pt minus 3pt}{1pt plus 1pt minus 1pt}
\titlespacing\subsection{0pt}{10pt plus 3pt minus 3pt}{1pt plus 1pt minus 1pt}
\titlespacing\subsubsection{0pt}{8pt plus 3pt minus 3pt}{1pt plus 1pt minus 1pt}

\providecommand{\keywords}[1]
{
  \small	
  \textbf{\textit{Keywords---}} #1
}

\usepackage{tikz,xcolor,hyperref}

\definecolor{lime}{HTML}{A6CE39}
\DeclareRobustCommand{\orcidicon}{
	\begin{tikzpicture}
	\draw[lime, fill=lime] (0,0) 
	circle [radius=0.16] 
	node[white] {{\fontfamily{qag}\selectfont \tiny ID}};
	\draw[white, fill=white] (-0.0625,0.095) 
	circle [radius=0.007];
	\end{tikzpicture}
	\hspace{-2mm}
}
\foreach \x in {A, ..., Z}{\expandafter\xdef\csname orcid\x\endcsname{\noexpand\href{https://orcid.org/\csname orcidauthor\x\endcsname}
			{\noexpand\orcidicon}}
}

\title{Efficient Detection of  Botnet Traffic by features selection and Decision Trees}

\usepackage{xwatermark}
\newwatermark[firstpage,color=gray!90,angle=0,scale=0.28, xpos=0in,ypos=-5in]{*correspondence: \texttt{javier.velasco@ unileon.es}}
\newwatermark[firstpage,color=black,angle=0,scale=0.32, xpos=0in,ypos=-4.5in]{Preprint submitted to IEEE Access}

\usepackage{authblk}

\author[1,2\thanks{\tt{javier.velasco@ unileon.es}}]{Velasco-Mata, Javier\orcidA{}}
\author[1,2]{Gonz\'alez-Castro, V\'ictor\orcidB{}}

\author[1,2]{\\Fidalgo, Eduardo\orcidC{}}
\author[1,2]{Alegre, Enrique\orcidD{}}

\affil[1]{Department of Electrical, Systems and Automation Engineering, Universidad de Le\'on}
\affil[2]{Researcher at INCIBE, Le\'on (Spain)}

\begin{document}

\twocolumn[ 
  \begin{@twocolumnfalse} 

\date{}
\maketitle

\begin{abstract}
Botnets are one of the online threats with the biggest presence, causing billionaire losses to global economies. Nowadays, the increasing number of devices connected to the Internet makes it necessary to analyze large amounts of network traffic data. In this work, we focus on increasing the performance on botnet traffic classification by selecting those features that further increase the detection rate. For this purpose we use two feature selection techniques, Information Gain and Gini Importance, which led to three pre-selected subsets of five, six and seven features. Then, we evaluate the three feature subsets along with three models, Decision Tree, Random Forest and k-Nearest Neighbors. To test the performance of the three feature vectors and the three models we generate two datasets based on the CTU-13 dataset, namely QB-CTU13 and EQB-CTU13. We measure the performance as the macro averaged F1 score over the computational time required to classify a sample. The results show that the highest performance is achieved by Decision Trees using a five feature set which obtained a mean F1 score of $85\%$ classifying each sample in an average time of $0.78$ microseconds.
\end{abstract}

\keywords{Computational and Artificial Intelligence, Computer Applications, Decision Trees, Machine Learning Algorithms, Optimization Methods} 

\vspace{0.35cm}

  \end{@twocolumnfalse} 
] 



\section{Introduction}
\label{sec:introduction}
One of the main current cyber-threats are botnets: networks of compromised and remote-controlled devices. These networks are dangerous because they can perform massive cyber-attacks, from  Distributed Denial-of-Service (DDoS) or spam campaigns \cite{zhao2013botnet} to disrupting critical structures such as power distribution grids \cite{wang2019securing}. In 2018, the average annual cost for a company due to incidents related to botnets was \$390,752 \cite{bissell2019cost}, and in 2019 the FBI's Internet Crime Complaint Center (IC3) recorded more than \$3.5 B in individual and companies losses related to cybercrime \cite{fbi2020icr}. Moreover, the ENISA report on botnets of 2020 \cite{ENISA2020} counted that 7.7 million of IoT devices are connected everyday to the Internet, increasing the surface attack for malware infections \cite{koroniotis2019forensics, li2019analysis}. Besides, it informed on an increase on botnet controlling servers of $71.5\%$ from the previous year, which brings up the need for developing countermeasures against botnets.

Traditionally, network traffic generated by malware has been detected with signature-based solutions such as Snort \cite{roesch1999snort} or Suricata \cite{albin2012realistic}. Although these methods are highly efficient once a proper malware-trait is obtained, they have several downsides. The first problem is that it requires a previous identification of the malware program \cite{aydin2009hybrid}. Second, it is necessary to perform a deep analysis of the malicious code to obtain the signature, which is also time-consuming \cite{liao2013intrusion}. Third, recent malware programs have started to use encryption in their communications \cite{anderson2016identifying}. This made the signatures based on Deep Packet Inspection (DPI) unusable, which are the ones that rely on inspecting the packets' payloads to determine if certain string patterns can be observed.

The disadvantages of signature-based methods can be compensated with behavior-based detectors, i.e., algorithms that classify traffic according to communication patterns \cite{garcia2009anomaly}. In our previous work \cite{velasco2019botnet}, we focused on determining the basis of a behavior-based botnet detector by creating a new botnet dataset and using it to measure the performance of four classifiers: Decision Tree (DT), k-Nearest Neighbors (k-NN), Support Vector Machine (SVM) and Na\"ive Bayes (NB). We used a selection of 13 traffic features and concluded that DT and k-NN achieved the highest detection scores. Besides, recent works also confirm the advantage of DT based algorithms to classify network traffic, as well as k-NN \cite{stevanovic2014efficient, khan2019adaptive, lee2021classification, alshamkhany2020botnet, jimenez2018effect}. For this reason, in this work, we selected DT, Random Forest (RF) and k-NN as candidate models to obtain an optimal botnet detection scheme.

While we found multiple academic works that employ a feature selection to improve the detection scores of the classifiers \cite{guerra2019hybrid, guerra2019towards, silva2020study, beigi2014towards, muhammad2020robust, koli2017advanced, beer2017feature, letteri2019feature, alejandre2017feature, jimenez2018effect}, we noticed a lack of research on the performance of the detectors after being trained, i.e., on the effects on the speed of the model at classifying samples. 
To give insight in this area, we took the feature set we used in \cite{velasco2019botnet} and determined the most relevant features. In this process we used Gini Importance and Information Gain, and obtained three feature subsets. Then, we contributed to this research line by measuring the performance of these three subsets. Note that by \textit{performance}, we refer to the relation between the achieved F1 and the computational time required by the feature subsets.

\begin{figure*}
\centering
\includegraphics[width=0.75\linewidth]{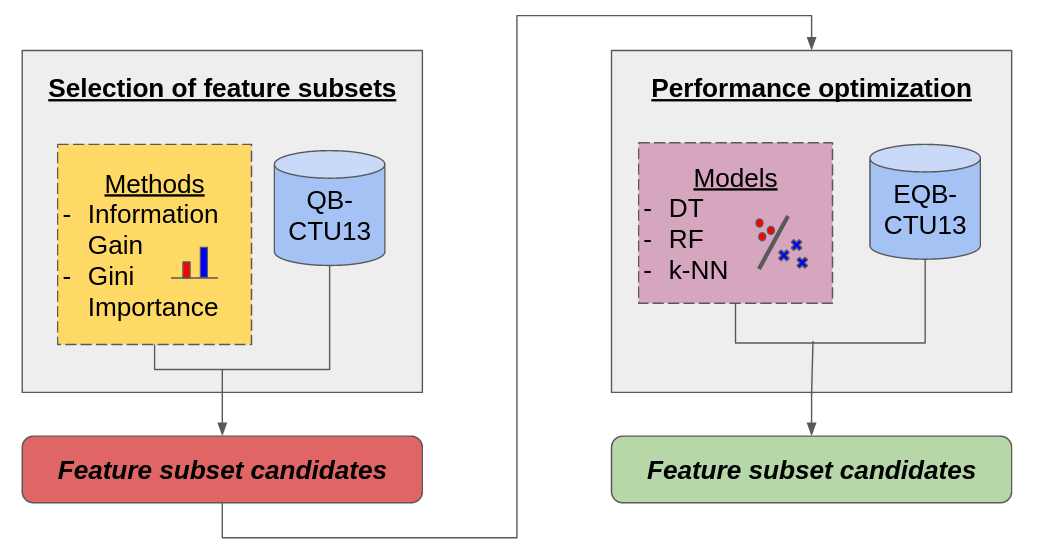}
\caption{Visualization of the experimentation.}
\label{fig:graph_abstr}
\end{figure*}

The other contribution we present in this work is developing two datasets based on the well-known CTU-13 database \cite{garcia2014empirical}. The main problem of CTU-13 is its imbalance between its classes, so we propose the Quasi-Balanced-CTU13 (QB-CTU13) as a selection of samples from the different classes while preserving the less represented ones, and adding traffic from regular sources (not botnets) to make a more updated normal class. Additionally, we included traces of three modern and hard to detect botnets to expand this dataset, creating the Extended QB-CTU13 (EQB-CTU13). The details of the datasets are explained in subsection \ref{subsec:datasets}. The graphical abstract of Fig. \ref{fig:graph_abstr} shows how these datasets are used along with the experimentation.

The structure of this work is described as follows. In Section \ref{sec:background} we review the literature related to botnet detection and feature selection. In Section \ref{sec:methodology} the methods used are described, while our experiments are detailed in Section \ref{sec:experimentation}. Finally, the results are presented and discussed on Section \ref{sec:results}, and the conclusions are drawn on Section \ref{sec:conclusions}.

\section{Background}\label{sec:background}

\subsection{Botnet detection}\label{background:botnet-detection}

Early methods for botnet detection such as BotMiner \cite{gu2008botminer} or BotSniffer \cite{gu2008botsniffer} relied on statistical algorithms to detect botnet traffic. This was possible by taking advantage of the spatial-temporal correlations of the traffic generated by the malware programs, as opposed to the more aleatory traffic produced by real humans. 

Reviewing the last publications on botnet detection, it can be observed that Machine Learning (ML) algorithms are the most common to build multi-class traffic classifiers. Among all of them, models based on Decision Trees (DT) usually achieve the best results. For example, in the work of Gadelrab et al. \cite{gadelrab2018botcap} Support Vector Machine (SVM) and DT were compared to select the best classifier for a botnet detector implementation (BoTCap). DT achieved the highest scores (F1 score of $95\%$) over a self-constructed dataset with traces of six botnets: Aryan, Ngr, Rxbot, Blackenergy, Zeus and Vertexnet. Moreover, DT based algorithms also showed high performance on network attacks datasets, like the CICIDS2017 dataset \cite{panigrahi2018detailed}, which collects online attacks such as DDoS or malware infections. This dataset was used in the work of MacKay et al. \cite{mckay2019machine} to construct three training sets and two testing sets, employed to compare six algorithms: DT, Random Forest (RF), k-Nearest Neighbors (k-NN), Na\"ive Bayes (NB), Multi-Layer Perceptron (MLP) and One Rule (OneR). The results showed that DT achieved the best F1 score ($99\%$) using the $78$ features of the CICIDS2017 dataset, and the authors left a feature selection as a future work.

Another related field where DT based models are used is detecting botnet accounts, i.e., automated users on social networks that propagate spam. In this problem, Chu et al. \cite{chu2012detecting} used an RF to classify a collection of 500.000 accounts of Twitter into three classes: human, bot and cyborg. The model obtained a True Detection Rate of $96\%$, where \textit{human} was the least misclassified class with only $1.4\%$ of human accounts wrongly classified. Machine Learning algorithms have also been used in other aspects to detect bot activity, such as over the logs of SSH sessions to identify malicious use of commands \cite{garre2021novel}. 

On the other hand, Deep Learning (DL) based classifiers were recently proposed for binary classification in botnet detection. For example, the work of Maeda et al. \cite{maeda2019botnet} used a Deep Neural Network (DNN) to identify the botnet traffic traces that reported an accuracy of $99.2\%$ in a dataset built by joining data from the CTU-13 dataset \cite{garcia2014empirical}, the ISOT dataset \cite{saad2011detecting} and self-captured not-malicious traffic. One of the downsides of ML and DL based detectors is the possibility of adversarial attacks, when the malicious program output is specially designed to mimic the features of a non-malicious one \cite{demontis2019yes}. Therefore, another benefit of DL approaches is the use of Generative Adversarial Networks (GAN), which generate 'fake' samples to be mixed within the dataset so the trained detector is more resilient against crafted attacks. Yin et al. \cite{yin2018enhancing} proposed a framework to improve botnet detectors by using a GAN called Bot-GAN. They used the ISCX 2014 dataset \cite{biglar2014towards} divided into $491,381$ samples for training and $348,452$ samples for testing, and they found that while the F1 score of the detector was $68.51\%$ when no fake samples where used, they achieved a peak F1 score of $70.59\%$ if $500$ fake samples were mixed with the training data.

To the best of our knowledge, the most used dataset to test botnet detectors is the CTU-13, either pure or with modifications. The CTU-13 is a public dataset created in the Czech Technical University (CTU) \cite{garcia2014empirical}, which contains data from seven different botnets: Neris, Rbot, Virut, Murlo, NSIS, Donbot and Sogou. Besides, the CTU-13 can either be used \textit{as a whole} or \textit{by scenario}. In total, the CTU-13 dataset contains 13 different scenarios, i.e., traffic captures in different situations where the monitored botnet performs one or more actions such as communication between bots, generating SPAM, or performing a DDoS attack.

The current purpose of the CTU-13 dataset is not to improve the detection rate of a previous work. In the recent literature, we found that the minimal accuracy obtained in all the scenarios of the CTU-13 dataset is $96.43\%$, achieved with a Self Organizing Map (SOM)\cite{le2019unsupervised}. Therefore, the current main utility of this dataset is to verify the detection ability of a solution. 

Other botnet types can be used to test if a traffic classifier can improve a previous result. We found that in the work of Abraham et al. \cite{abraham2018comparison} the best F1 score obtained for the botnet Bunitu was $90\%$ with an ensemble of classifiers. Moreover, the work of AlAhmadi \& Martinovic \cite{alahmadi2018malclassifier} reported that the recall obtained for the NotPetya botnet using an RF classifier was $60\%$, and that $25\%$ of the samples of this botnet were wrongly classified as produced by the botnet Miuref.

\subsection{Feature selection}

In general, feature selection techniques can be categorized as wrapper, filter and embedded methods \cite{li2018feature}.

Wrapper methods search the most optimal feature subset for a given learning algorithm through iteration. Their main downside is that the found feature subset is designed for the selected learning algorithm, and it is not guaranteed to be optimized for any different algorithm \cite{wang2015accelerating}. 

Filter methods overcome this problem by not evaluating the feature subset with a learning algorithm. Instead, they rely on data characteristics such as entropy or feature correlations. These methods are usually more computationally efficient than wrapper methods, but the obtained feature subset may not be the most optimal for a given learning algorithm \cite{ambusaidi2016building}.

Finally, the embedded methods suppose a mixture of the previous two ones. The performances of the feature subsets are evaluated with a pre-selected learning algorithm, but it does not use an iterative search as wrapper methods. Instead, they assign weights to the initial features and try to minimize (or nullify) the value of the weights, which represent the feature importance \cite{jovic2015review}.

In the context of feature selection for botnet detection, different methods have been applied. For example, the work of Gadelrab et al. \cite{gadelrab2018botcap} used the Weka framework to implement an embedded feature selector using "BestFirst" as search algorithm and "CfsSubsetEval" as evaluating method. This allowed the creation of feature subsets which afterwards were evaluated with a DT classifier and three SVM classifiers. The evaluated SVM differed in the selected kernel function: linear, polynomial of Radial Basis Function (RBF). The best result, with an F1 score of $95\%$, was achieved with a subset of five features and the DT classifier. The dataset used was constructed by the authors and was limited to IRC and HTTP based botnets.

Alauthaman et al. \cite{alauthaman2018p2p} tested three feature reduction methods: entropy impurity, RelieF and Principal Component Analysis (PCA). The target classification algorithm was a Neural Network (NN), and the most efficient result was obtained using the entropy impurity based feature selection. The achieved accuracy  was $99.20\%$ after reducing the feature set from $29$ to $10$ features. The data used was a mixture of traffic traces from the ISOT \cite{saad2011detecting} and the ISCX \cite{shiravi2012toward} datasets. 

Moreover, Ferhat O. Catak \cite{catak2018two} compared four classifiers for malware traffic detection: DT, RF, NN, AdaBoost and NB. They used a filter feature selection based on the L1-norm to reduce the feature pool from $36$ to $11$ features, achieving the highest accuracy with RF ($89.04\%$) followed by DT ($86.77\%$). The dataset used was created by the Australian Centre for Cyber Security (ACCS) \cite{moustafa2015unsw}.

The work of Letteri et al. \cite{letteri2019feature} focused on improving the F1 score through feature reduction on a NN with two feature selection methods: a Gini Importance derived selection, and Information Gain (also known as Mutual Information). The features were based on time windows, by counting the number and type of packets sent to a certain IP inside a time interval. The best results achieved a F1 score of $97.30\%$ with both the Gini Importance and Information  Gain based selections. The dataset used was generated from traffic captures offered by the Stratosphere IPS \cite{stratosphereips}.

Feng et al. \cite{feng2018feature} used PCA and an embedded method based on SVM to determine the importance of an initial set of $55$ features for cyber attacks detection. The selected datasets were the C08, C09 and C13 ones from the CCC dataset collection \cite{mws2016}, and the feature subsets were evaluated using SVM and RF. They determined the top $10$ most important features and concluded that they could reduce the total set of $55$ features to $40$ without experiencing a significant decrease in the detection rate. After this work, Muhammad et al. \cite{muhammad2020robust} used the CCC dataset collection and also selected a subset of $40$ features but using PCA and Information Gain for feature selection. They tested more models, RF, SVM, Logistic Regression (LR) and MLP and were able to enhance the result of Feng et al. using RF and obtaining a $99\%$ as the highest accuracy.

In addition, DL methods can also be used in feature reduction and selection.  For example, the work of Parker at al. \cite{parker2019demise} used an AutoEncoder (AE) Neural Network to obtain a set of $50$  new features derived from the original set of $154$ features. They combined the two sets and ranked the importance of the $204$ features with the Information Gain method. To conclude, they determined that the most optimal solution was an LR classifier using five features, obtaining an F1 score of $98.01\%$ with a computational training  cost of $603.33$ seconds on a computer with a $3.1$ GHz Intel Core i7 CPU and 8 GB of RAM. The dataset used was the reduced CLS portion of the AWID dataset \cite{kolias2015intrusion}. 

Finally, State-of-the-Art works show that it is possible to implement botnet detectors using a hyper-reduced set of features. In 2020, Joshi et al. \cite{joshi2020analysis} used five feature selection techniques to reduce their original set of features, and a reduced feature set was proposed from each method. The smallest set proposal came from the output of PCA, which had only two features: the source IP and the protocol. Besides, they tested the feature subsets with four classifiers, SVM, LR, k-NN and DT. Using the CTU-13 dataset, k-NN gave the best results: a $0.99$ of accuracy using only the two features set. A year later, Ismail et al. \cite{ismail2021effects} conducted a similar research using more recent data. They used seven traffic captures from the Stratosphere IPS \cite{stratosphereips}, and they also tested more classifiers: Bayesian Network, NB, k-NN, Ada Boost, SVM, J48, RF, Ripper and PART. They concluded that SVM attained the best true positive rate on detecting botnet data across the seven traffic captures, using a set of only three features: the destination and source IPs and the protocol. Despite these works providing insights on reduced feature sets, they did not perform an analysis on the computational cost of their proposals.

\section{Methodology}\label{sec:methodology}

For selecting a feature subset capable of maintaining a good detection rate and reducing the classifying time of the ML models, we elaborated the following methodology. First, as recommended by \cite{guerra2019hybrid} we used a wrapper method to select the three features that contribute the most to the F1 score. One common wrapper strategy usually followed is to rank the importance of each feature, and then to check the evolution of the F1 score when adding the features one by one according to the ranking \cite{silva2020study, alejandre2017feature,  beigi2014towards}. Sometimes, this strategy can be improved by ranking the features with two different methods, as it was done in \cite{muhammad2020robust} where they used both PCA and Information Gain to rank the features they selected to classify botnet traffic. 

However, it is also known that using PCA on supervised learning can be counterproductive \cite{PCA2019}. Because of that,  we used Gini Importance instead of PCA, since it measures how well can the data be divided into homogeneous groups, which is important for the three classifiers used in our experiments. Finally, after selecting the three subsets with the features that contribute the most to the F1 score, we measured the trade-off between the F1 score and the computational cost, in terms of the mean time needed to classify a sample, of the trained models using each of these subsets.

\subsection{Feature ranking methods}

Before selecting a reduced feature subset it is necessary to determine the importance of each feature. As we just justified above, we used concurrently two methods to rank features: Gini Importance and Information Gain.

\subsubsection{Gini Importance}

The Gini Gain is defined as the variation of the Gini Impurity after a split of the data using a feature \cite{nembrini2018revival}. This metric is perceived as more \textit{important} as the decrease in the Gini Impurity is higher, and thus we can rank the features according to this \textit{Gini Importance}.

The Gini Impurity of a subset of samples $\Gamma$ is defined in \eqref{eq:gini_impurity_subset}, where $p(i)$ is the probability to pick a subset sample of class $i$, and $J$ is the number of classes present in the subset.

\begin{equation}
\label{eq:gini_impurity_subset}
\Gamma=\sum_{i=1}^{J} p(i)*(1-p(i))
\end{equation}

Finally, the Gini Impurity of the whole dataset $\Gamma_X$ in a given state, before or after one or more splits, is the weighted mean of the Gini Impurity of its subsets $\Gamma_j$ as shown in \eqref{eq:gini_impurity}, where $w_j$ represents the $j$ subset weight.

\begin{equation}
\label{eq:gini_impurity}
\Gamma_X=\frac{1}{n} \sum_{j=1}^{n} w_j\Gamma_j
\end{equation}

\subsubsection{Information Gain}

Information Gain allows us to quantify which feature provides maximal information about the classification, using for that the notion of entropy.
The Information Gain obtained with a feature is defined as reducing the entropy in the dataset after knowing the values of the samples for this feature \cite{wen2016accurate}.

On the one hand, the original entropy of the dataset $H(X)$ is defined in \eqref{eq:original_entropy} based on the probability $p(x)$ of a sample to belong to the class $x$. On the other hand, the entropy $H(X|Y)$ of the dataset after knowing the values of the feature $Y$ is defined in \eqref{eq:reduced_entropy} based on the probability $p(y)$ of a sample to present the value $y \in Y$, and the probability $p(x|y)$ of a sample of class $x$ to present the feature value $y \in Y$.

\begin{equation}
\label{eq:original_entropy}
H(X)=- \sum_{x=1}^{X} p(x) log (p(x))
\end{equation}

\begin{equation}
\label{eq:reduced_entropy}
H(X|Y)= - \sum_{y} p(y) \sum_{x}^{X} p(x|y) log (p(x|y))
\end{equation}

\subsection{Feature vector evaluation} \label{subsec:feature_vector}

For this work we only considered traffic based on the TCP protocol because it is used by the majority of known botnets \cite{wainwright2019analysis}.
Previously we used thirteen features to characterize TCP flows \cite{velasco2019botnet}, but we dismissed two of them in this work: the average and the variance of the packets' velocity transmission  (mVel and vVel). We removed them because their information was already represented by the average and variance on the interval time between consecutive packets (mTime and vTime), as shown in Equation \eqref{eq:redundancy}. Table \ref{tab:original-features} shows the remaining eleven features we selected to model network traffic.

\begin{equation}
\label{eq:redundancy}
\begin{split}
\textrm{mVel} & =\frac{\textrm{number of packets in TCP flow}}{\textrm{duration of TCP flow}} \\
              & =\frac{1}{\textrm{mTime}}
\end{split}
\end{equation}

\begin{table}[!t]
\renewcommand{\arraystretch}{1.3}
\caption{Network Traffic Features \cite{velasco2019botnet}}
\label{tab:original-features}
\centering
\begin{tabular}{l p{5.1cm}}
Feature(s) & Description \\
\hline
sPort, dPort  & Source and destination ports in the TCP connection \\

mLen, vLen  & Mean and variance of the payload lengths of all packets in the TCP connection \\

mTime, vTime & Mean and variance of the interval times between sent packets in the TCP connection \\

mResp, vResp & Mean and variance of the interval times between a packet reaches the machine that started the connection and it responds to that packet \\

nBytes & Total number of bytes exchanged in the TCP connection \\

nSYN & Total number of SYN flags exchanged during the TCP connection \\

nPackets & Total number of packets exchanged in the TCP connection \\

\end{tabular}
\end{table}

We used three measures to evaluate different subsets of features: (1) the F1 score to measure the capability to detect botnet samples, (2) the computational time to consider the cost of the classification of a sample, and (3) the performance expressed as a relation between the former two. 

The F1 score is defined as the harmonic mean of the precision and the recall \cite{sasaki2007truth}. In the context of this work, the precision is the fraction of samples detected as ``botnet'' that are actually generated by malware; while the recall is the fraction of botnet samples that are detected, over all the botnet samples in the dataset \cite{mccallum1998comparison}.

The computational time is the total time required by a computer with a specific processor to conclude a task. In this work, we used this time to measure the computational cost of classifying a sample, considering that all the experiments were made with the same CPU. The measured time did not include either the training time of the model or the time required to calculate the traffic features since they were already present in the datasets.

To measure the performance of a vector of $n$ features, we use the ratio between the F1 score and the computational time, which measures the gain in detection capability by the cost of this detection.

\subsection{Classifiers}
\label{subsec:classifiers}

In this work, we carry out our experiments with three Machine Learning classifiers: Decision Tree (DT), Random Forest (RF) and k-Nearest Neighbors (k-NN).

The DT can be visualized as a tree-structured set of questions or decisions about the sample to be classified \cite{buntine1992further}. Depending on the values of the sample features, a path in the tree of decisions will be followed to estimate the class of the sample. 

The RF classifier is a conglomerate of $m$ DT classifiers \cite{liaw2002classification}. In the training phase, each DT is trained with a bootstrap sample \cite{singh2008bootstrap} of the training data; and in the classification process, the output is generated with a voting system among the DT models.

The third algorithm is k-NN. If the samples are characterized with $m$ features, in the training phase the k-NN classifier stores the samples in a $m$ dimensional hyperspace. In the classification phase, each sample is projected in that hyperspace and the algorithm searches for the $k$ nearest training samples to determine its class \cite{dong2011efficient}.

\section{Experimentation}\label{sec:experimentation}

\subsection{Datasets}\label{subsec:datasets}

The experimentation was made using two datasets \textit{QB-CTU13} and \textit{EQB-CTU13}, which we built based on the CTU-13 dataset \cite{garcia2014empirical} to overcome its class imbalance. The significant differences in the number of usable TCP flows of the CTU-13 dataset may suppose an issue in the multi-classification since the algorithms tend to over-fit the most represented classes. The least represented botnet class (Sogou) has only $36$ usable TCP flows so balancing the dataset to this class would mean a significant waste of data. Not considering the smaller classes would reduce the dataset to hold only data from three botnets.

The \textit{Quasi-Balanced based on CTU-13 dataset} (QB-CTU13) contains the same number of usable TCP flows from the three most represented botnets and all the usable TCP flows from the remaining ones, hence the name `Quasi-Balanced'. We developed this dataset because it allowed us to obtain a more meaningful weighted F1 score by avoiding that the results of the most represented classes would hide the ones of the less represented classes. 

\begin{figure}
\centering
\includegraphics[width=\linewidth]{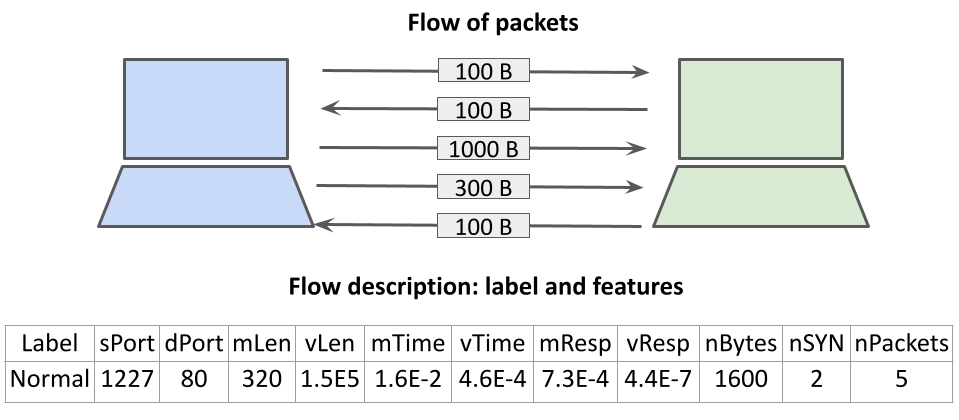}
\caption{Two computers establish a communication using the TCP protocol. The set of interchanged packets is named flow, and by extracting its features we constituted a sample.}
\label{fig:get_sample}
\end{figure}

The \textit{Extended QB-CTU13} (EQB-CTU13) is constructed using the same TCP flows from the QB-CTU13, and TCP flows from the Stratosphere IPS \cite{stratosphereips} of three new botnet classes: Bunitu, Miuref and NotPetya. We used all the available samples from these three botnets in their repository by the end of 2020. We selected these new particular botnets to be able to compare our study with the works of Abraham et al. \cite{abraham2018comparison} and AlAhmadi \& Martinovic \cite{alahmadi2018malclassifier}. The number of used TCP flows of each class in both datasets is specified in Table \ref{tab:dataset}. We obtained a sample from each flow as pictured in Fig. \ref{fig:get_sample}, where we also show the features we extracted.

\begin{table}[!t]
\renewcommand{\arraystretch}{1.3}
\caption{TCP flows used as Class Samples in QB-CTU13 and EQB-CTU13 Datasets.}
\label{tab:dataset}
\centering
\begin{tabular}{c c c}
Class & QB-CTU13 & EQB-CTU13 \\ 
\hline
Normal & 3890 & 3890\\
Neris & 3890 & 3890\\
Rbot & 3890 & 3890\\
Virut & 3890 & 3890\\
Murlo & 2036 & 2036\\
NSIS & 355 & 355\\
Donbot & 233 & 233\\
Sogou & 36 & 36\\
\textbf{Bunitu} & - & 3890 \\
\textbf{Miuref} & - & 3890 \\
\textbf{NotPetya} & - & 111 \\
\end{tabular}
\end{table}

\subsection{Feature evaluation and feature subset selection}

The experimentation completed in this work was accomplished on a machine with 128 GB of RAM and two Intel(R) Xeon(R) E5-2630v3 CPUs running an Ubuntu 14.04.5 LTS. The software was developed using Python 3 and the library scikit-learn\footnote{http://scikit-learn.org/stable/index.html}.

As we already explained, to evaluate the information provided by each of the eleven features described in subsection \ref{subsec:feature_vector}, we used two filter selection techniques, Gini Importance (GI) and Information Gain (IG), to rank them over the QB-CTU13 dataset from these two different perspectives. Next, we analyzed the evolution of the F1 score over the same dataset when introducing a new feature, increasing the number $n$ of used features, from $n=1$ to $n=11$. We experimented with the two different ways of adding features, both following the GI and IG rankings, and we obtained the F1-score with DT, RF and k-NN. In each of these tests, the F1 score was calculated as the weighted average of the results of a 10-fold cross-validation.

The DT was optimized by not limiting the number of generated branches, while we searched for the most optimal versions of RF and k-NN by using a different number, $m$, of DTs ($m\in\{10,20,...,100\}$) and different amount, $k$, of neighbors ($k\in\{1,3,5,7,9\}$) respectively. 

By observing the evolution of the F1 score, we built three feature subsets containing the $n$, $n-1$ and $n-2$ features that produce the most notable increases in the F1 score among the three classifiers. Finally, we used the EQB-CTU13 dataset to study the performance of the three optimized classifiers using each one of the three feature subsets. Since the F1 score is calculated as the weighted average of the results of a 10-fold cross-validation, for calculating the computational time we used a model trained with the same proportion of the dataset, a $90\%$ of the samples, and we used the rest of the dataset for calculating the time required for classifying the samples.

\section{Results and discussion}\label{sec:results}

\begin{figure*}
\centering
\includegraphics[width=0.75\linewidth]{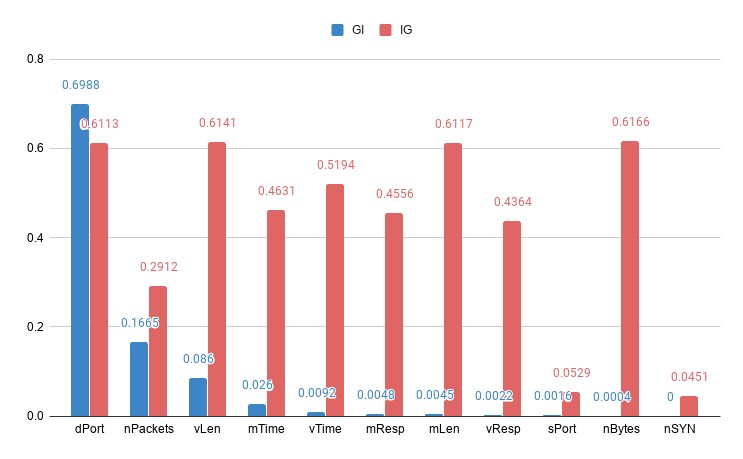}
\caption{Gini Importance (GI) and Information Gain (IG) of the 11 selected features over the QB-CTU13 dataset.}
\label{fig:ig-gi}
\end{figure*}

As we pointed out before, we analyzed the importance of the eleven selected features with two feature selection methods, Gini Importance (GI) and Information Gain (IG), over the QB-CTU13 dataset, as shown in Fig. \ref{fig:ig-gi} 
. We observed the disparity between the source (sPort) and destination (dPort) ports. We assumed this fact was reasonable since the dPort is related to the type of service used in the connection, e.g., the port 80 is normally linked to HTTP traffic. However, the sPort can be changed by a Network Access Translation (NAT) system or auto-assigned by the Operating System without the intervention of the program that originated the connection. Besides, we observed that both selection methods consider the number of exchanged SYN flags (nSYNflags) as the least important feature.

\begin{figure*}
\centering
\includegraphics[width=0.75\linewidth]{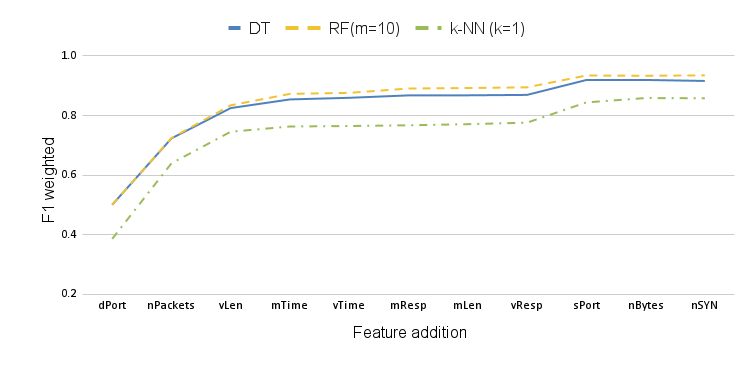}
\caption{Evolution of the F1 scores by increasing the number of considered features following the Gini Importance ranking. The $x$ axis shows the feature that is added: First, the models started only using dPort, then they used dPort and nPackets, and so on. In the case of RF and k-NN, only the optimal values of $m$ and $k$ are represented.}
\label{fig:gini_all_complete}
\end{figure*}

\begin{figure*}
\centering
\includegraphics[width=0.75\linewidth]{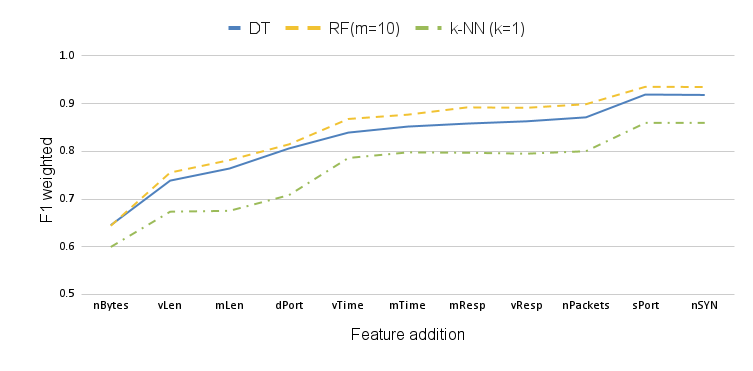}
\caption{Evolution of the F1 scores by increasing the number of considered features following the Information Gain ranking. The $x$ axis shows the feature that is added: First, the models started only using nBytes, then they used nBytes and vLen, and so on. In the case of RF and k-NN, only the optimal values of $m$ and $k$ are represented.}
\label{fig:ig_all_complete}
\end{figure*}

Once we obtained the two feature rankings shown in 
Fig. \ref{fig:ig-gi}
, we measured the evolution of the weighted F1 score when an increasing number of features is considered. 
We added a feature each time, one by one, from the most important (GI), see Fig. \ref{fig:gini_all_complete},  or most informative (IG), see Fig. \ref{fig:ig_all_complete}, to the least relevant.

\begin{figure*}
\centering
\includegraphics[width=\linewidth]{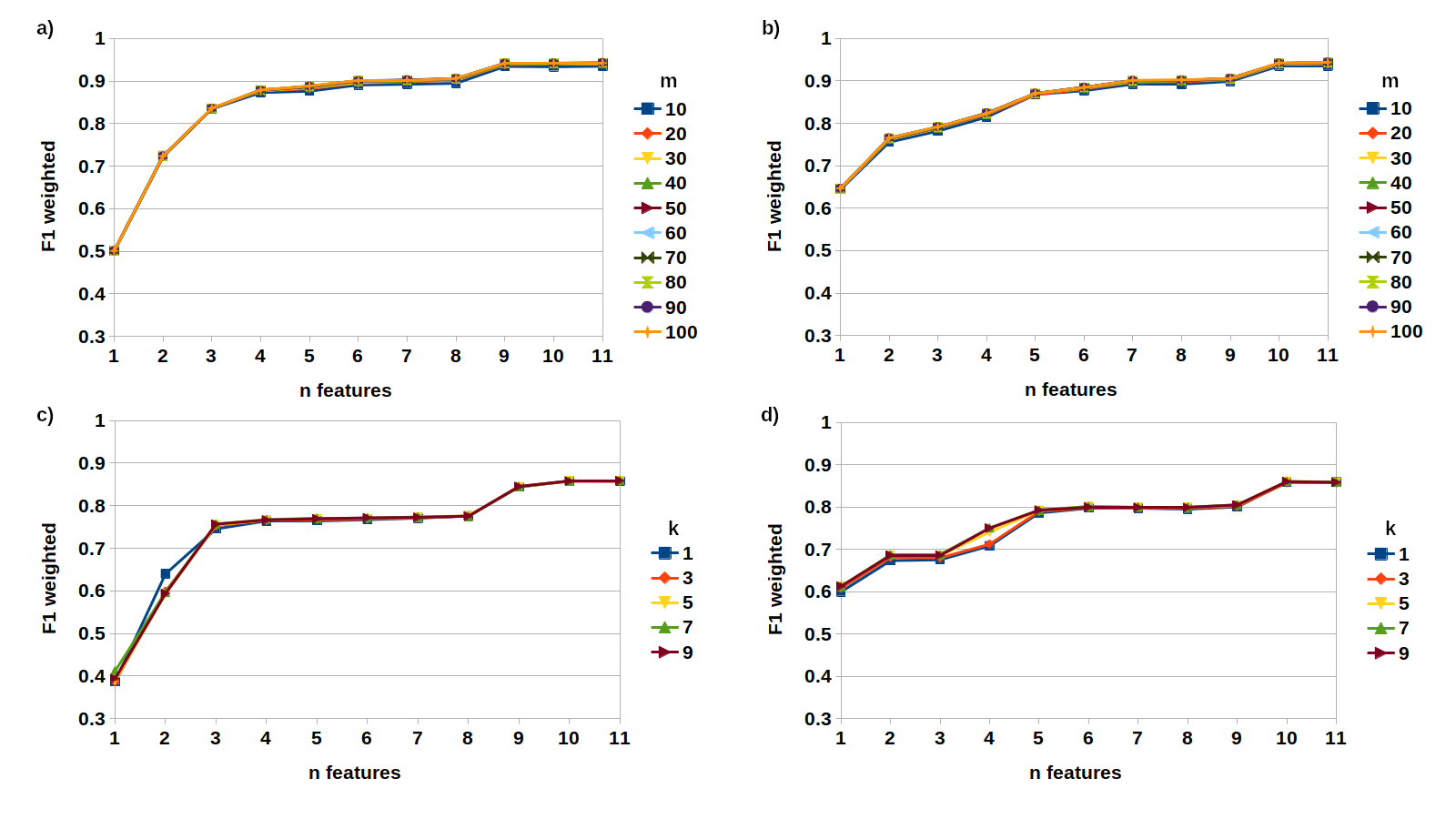}
\caption{Evolution of the F1 score by adding features to classify the traffic. On the top graphics (a and b) we compare the F1 score of different versions of RF integrating $m$ DTs, and on the bottom (c and d) we compare different versions of k-NN by varying the number $k$ of neighbors. On the left graphics (a and c), the features were added following the Gini Importance, while on the right ones (b and d) they were added following the Information Gain.}
\label{fig:kandm}
\end{figure*}

The F1 scores were calculated using DT, RF and k-NN,  looking for the most optimal versions in terms of performance of the RF and k-NN. In the RF case, we experimented with different numbers of DT ($m \in \{10,20,30,40,50,60,70,80,90,100\}$), and in the k-NN case we considered several numbers of neighbors ($k \in \{1,3,5,7,9\}$). Fig. \ref{fig:kandm} shows the F1 scores obtained with each version of the RF and the k-NN. It is observable that there was little effect by varying the values $m$ for RF and $k$ for k-NN. Thus, we selected the smaller values since they favored the speed of the classifiers. To compare RF and k-NN with DT, in Fig. \ref{fig:gini_all_complete} and Fig. \ref{fig:ig_all_complete} we only represented the results of the optimal values $m=10$ for RF and $k=1$ for k-NN, and also because it improved the data visualization.

Fig. \ref{fig:gini_all_complete} shows the gradual increase of the F1 score by adding features following the order dictated by the GI ranking. We observed that dPort, nPackets, vLen, mTime and sPort are, in this order, the features that generated the most significant increment of F1 score.

Similarly, when observing in Fig. \ref{fig:ig_all_complete}  the growth of  F1-score by adding features following the IG order, the features that caused the most significant F1 raises were nBytes, vLen, mLen, dPort, vTime and sPort.

As explained previously, the sPort might be hidden behind a NAT service, and therefore we did not consider this feature to be a candidate for the proposed optimal subsets. Therefore, as candidate features, we selected the ones that, by being added, produced the most significant increases on the F1 score in the previous experiments: dPort, vLen, nBytes, nPackets, mLen, mTime and vTime, making a total of $n=7$ features.

We selected three slightly different subsets of features to search for the optimal performance containing five, six and seven features each. The initial five-features subset is composed of the two most relevant features for each method of selection used, that in the case of GI are dPort and nPackets and for GI are nBytes and vLen. We also include mLen, which is the third one proposed by IG, because its information coefficient is very close to vLen, the second one. The fourth one in the IG list is dPort, already included since it is the first one in the GI list. We decided to exclude the fifth one, vTime, seeing that its coefficient was not so close to the previous four features and its relevance, in the GI list, is very small.

We decided to include mTime and vTime for the seven features subset, because they are the two following most relevant features in both lists. Moreover, to evaluate the performance with a smaller feature vector, we decided to drop vTime for the six-features vector due to its GI coefficient being very low.

Therefore, the three different feature vectors evaluated were the following ones:

\begin{enumerate}
\item Five feature vector: [dPort, nPackets, nBytes, vLen, mLen]
\item Six feature vector: [dPort, nPackets, nBytes, vLen, mLen, mTime]
\item Seven feature vector: [dPort, nPackets, nBytes, vLen, mLen, mTime, vTime]
\end{enumerate}

\begin{figure}
\centering
\includegraphics[width=\linewidth]{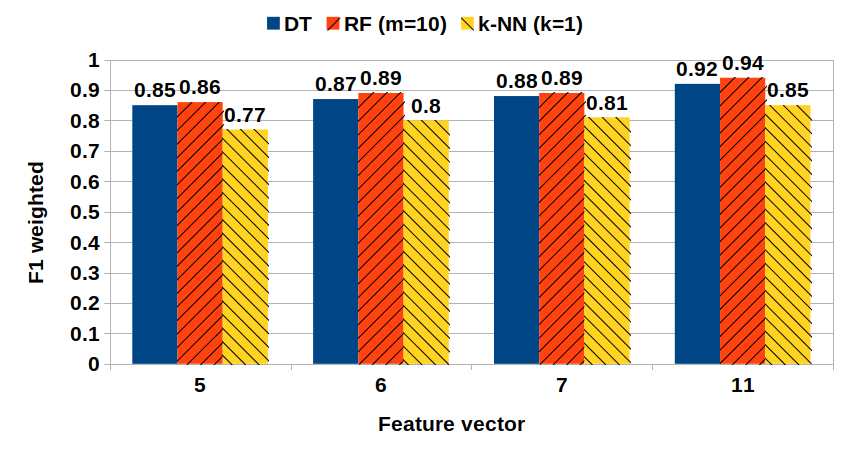}
\caption{F1 scores of DT,  RF ($m=10$) and k-NN ($k=1$) using the 5, 6 and 7-features subsets, as well as using the complete set of eleven features. The F1 used a weighted average among classes of the EQB-CTU13 dataset.}
\label{fig:f1_final}
\end{figure}

\begin{figure}
\centering
\includegraphics[width=\linewidth]{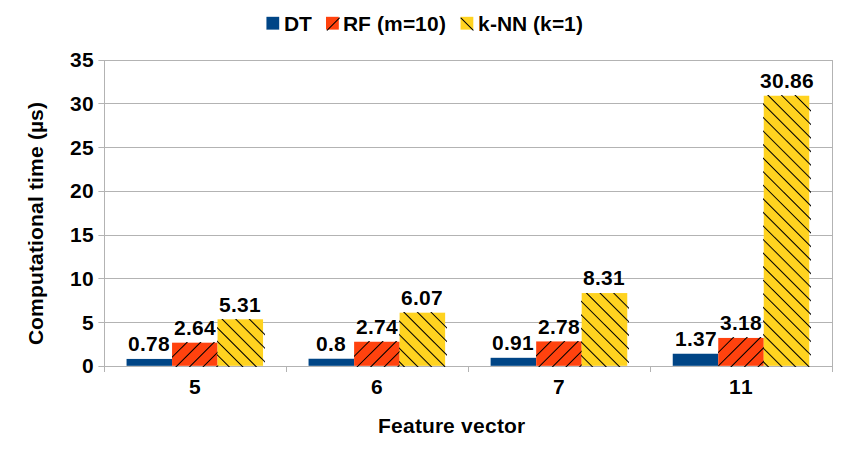}
\caption{Computational time to classify a sample needed by DT,  RF ($m=10$) and k-NN ($k=1$) using the 5, 6 and 7-features subsets and the complete set of eleven features over the EQB-CTU13 dataset.}
\label{fig:computation_final}
\end{figure}

\begin{figure}
\centering
\includegraphics[width=\linewidth]{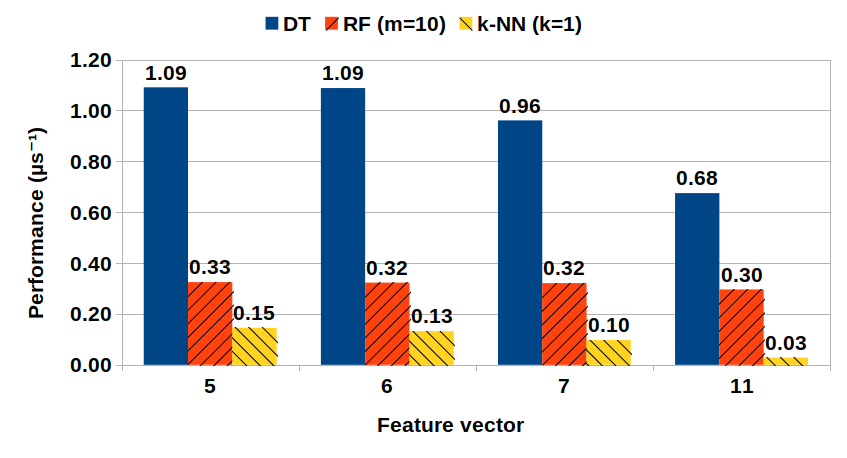}
\caption{Performance achieved by DT,  RF ($m=10$) and k-NN ($k=1$) using the 5, 6 and 7-features subsets and the complete set of eleven features over the EQB-CTU13 dataset.}
\label{fig:performance_final}
\end{figure}

These selected feature subsets were tested with the three optimized models, DT, RF ($m=10$) and k-NN ($k=1$) over the EQB-CTU13 dataset. Fig. \ref{fig:f1_final} represents the weighted F1 scores, and shows a clear advantage of DT and RF over k-NN. Moreover, considering the computational cost of the models shown in Fig. \ref{fig:computation_final}, k-NN also needed a longer time for classification than the other models. Besides, it is expected that it will perform even worse in terms of computational time with larger datasets, since it stores all the training data and searches for the $k$ most similar samples. While RF achieved higher F1 scores than DT in Fig. \ref{fig:f1_final}, the cost of using 10 DTs instead of one meant a significant reduction in the performance, as shown in Fig. \ref{fig:performance_final}. The tested model that achieved the best performance was DT using the 5-feature subset, and the DT using a 6-feature subset achieved a close result. However, using an extra feature also means that it has to be calculated. Thus under the perspective of performance, the DT with the 5-feature subset is more recommendable.

On Table \ref{tab:dt_by_class} can be seen the F1, recall, precision and performance values by each class of the EQB-CTU13 dataset achieved in a 10-fold cross-validation with two versions of the DT: using the performance-optimal 5-features subset, and using all the eleven features. The performance was defined as the ratio between the F1 score and the time required to classify a sample. We also compared our proposal with the SotA proposals of Joshi et al. \cite{joshi2020analysis}, a k-NN using two features (knn-2f); and Ismail et al. \cite{ismail2021effects}, a SVM using three features (svm-3f). According to the experimentation of these two works, k-NN and SVM obtained better results than DT using their proposed two and three feature sets respectively.

\begin{table*}[t]
\caption{F1, Recall, Precision and Performance Scores of our proposal, a DT with a 5-feature subset (dt-5fs). For comparison, we include the scores of a DT using the initial set of 11 features (dt-11fs), the SVM proposed by Joshi et al. \cite{joshi2020analysis} which uses three features (svm-3fs), and the k-NN proposal of Ismail et al. \cite{ismail2021effects} which uses two features (knn-2fs).}
\label{tab:dt_by_class}
\centering
\resizebox{\textwidth}{!}{%
\begin{tabular}{c c c c c c c c c c c c c c c c c}
 & \multicolumn{4}{c}{F1} &  \multicolumn{4}{c}{Recall}  &  \multicolumn{4}{c}{Precision} &  \multicolumn{4}{c}{Perf. (${ms}^{-1}$)} \\
Class & dt-5fs & dt-11fs & svm-3f & knn-2f & dt-5fs & dt-11fs & svm-3f & knn-2f & dt-5fs & dt-11fs & svm-3f & knn-2f & dt-5fs & dt-11fs & svm-3f & knn-2f \\
\hline
Normal  & 0.78 & 0.93 & 0.99 & 0.99 & 0.83 & 0.93 & 0.99 & 0.99 & 0.74 & 0.93 & 1.00 & 1.00 & \textbf{1007.49} & 68.32 & 2.16 & 48.51\\
Bunitu  & 0.84 & 0.94 & 0.01 & 0.99 & 0.87 & 0.94 & 0.07 & 0.99 & 0.81 & 0.94 & 0.01 & 0.99 & \textbf{1079.65} & 682.32 & 0.03 & 48.46 \\
Donbot  & 0.88 & 0.85 & 0.64 & 0.98 & 0.85 & 0.87 & 0.52 & 0.98 & 0.91 & 0.83 & 0.84 & 0.99 & \textbf{1139.20} & 620.69 & 1.38 & 47.97 \\
Miuref  & 0.99 & 1.00 & 0.37 & 1.00 & 0.99 & 1.00 & 0.23 & 1.00 & 0.99 & 1.00 & 1.00 & 1.00 & \textbf{1278.51} & 728.21 & 0.81 & 48.76 \\
Murlo   & 0.98 & 0.97 & 0.00 & 1.00 & 0.97 & 0.97 & 0.00 & 1.00 & 0.99 & 0.97 & 0.00 & 1.00 & \textbf{1259.58} & 709.21 & 0.00 & 48.70 \\
NSIS    & 0.59 & 0.85 & 0.00 & 0.01 & 0.59 & 0.86 & 0.00 & 0.11 & 0.59 & 0.84 & 0.00 & 0.10 & \textbf{761.88} & 623.66 & 0.00 & 0.29 \\
Neris   & 0.69 & 0.84 & 0.00 & 1.00 & 0.71 & 0.83 & 0.00 & 1.00 & 0.67 & 0.85 & 0.00 & 1.00 & \textbf{891.02} & 610.92 & 0.00 & 48.63 \\
NotPet. & 0.96 & 0.95 & 0.00 & 0.99 & 0.97 & 0.95 & 0.00 & 1.00 & 0.95 & 0.94 & 0.00 & 0.98 & \textbf{1241.69} & 693.49 & 0.00 & 48.28 \\
Rbot    & 0.98 & 0.99 & 0.50 & 0.96 & 0.98 & 0.99 & 0.58 & 0.92 & 0.98 & 0.99 & 0.45 & 0.99 & \textbf{1262.78} & 722.94 & 1.09 & 46.62 \\
Sogou   & 0.22 & 0.55 & 0.00 & 1.00 & 0.38 & 0.48 & 0.00 & 1.00 & 0.15 & 0.64 & 0.00 & 1.00 & 285.76 & \textbf{400.61} & 0.00 & 48.78 \\
Virut   & 0.75 & 0.84 & 0.32 & 1.00 & 0.78 & 0.85 & 0.56 & 1.00 & 0.72 & 0.83 & 0.22 & 1.00 & \textbf{967.83} & 614.89 & 0.69 & 47.71 \\
\end{tabular}
}
\end{table*}

Regarding the two versions based on DT, in three botnet classes, Donbot, Murlo and NotPetya, the F1 score achieved with five features is better than when using the eleven features. The recall score obtained for NotPetya is higher when using the 5-feature subset. Although the eleven feature vector achieves better detection results, the performance metric is significantly higher with the five feature vector, except for the class Sogou, which is explainable by the scarce representation of this class (only $36$ samples or TCP flows). This also caused the botnet Sogou to obtain the lowest scores in F1 and recall. Other works \cite{le2019unsupervised} obtained a higher score for this botnet in the CTU-13 dataset since they worked with each scenario separately. 

As we indicated before, to create the EQB-CTU13 dataset from QB-CTU13 we added traffic samples from the Bunitu, NotPetya and Miuref botnets in order to compare our models with previous results from other authors. In Abraham et al. \cite{abraham2018comparison}, the best F1 score achieved for the Bunitu botnet was $90\%$, while with our approach, using eleven features we obtain a F1 of $94\%$ and, with the most time-efficient subset, with five features, the F1 score achieved was $84\%$.

In relation to NotPetya and Miuref botnets, AlAhmadi \& Martinovic \cite{alahmadi2018malclassifier} obtained a recall for the NotPetya botnet of $60\%$, where $25\%$ of the NotPetya samples were wrongly classified as Miuref samples. In our proposal, the performance-optimized DT achieved  a recall of $96\%$ for the NotPetya class, and the F1 and recall scores of the Miuref class were both of $99\%$.

Finally, on one side, the k-NN using two features attained the best results in all the classes. These two features were the source IP and the protocol used. On the other side, the SVM also used these two features and added the destination IP, and yielded worse results. It is observable that the SVM using three features attains a good performance distinguishing between normal and botnet traffic, but it suffers in the task of differentiating botnets. However, the performance of these two classifiers was in the order of ${10}^{-2}$ lower than the performance of the DT, which indicated that even using more features and also considering the extra time to calculate them, the DT is significantly faster than SVM and k-NN.

\section{Conclusions and future works}\label{sec:conclusions}

In this work, we propose a reduced set of features to detect and classify, with a high accuracy, some of the most popular bot networks. Our proposal is more efficient in terms of time and obtains a similar or better accuracy than other approaches provided for classifying these botnets.

We selected the most relevant features using the Gini Importance and Information Gain criteria. With the selected features we proposed three subsets, evaluating its performance using Decission Trees (DT), Random Forest (RF) and K-Nearest Neighbors (k-NN) for botnet detection. We optimized RF and k-NN for botnet detection tuning their respective parameters $m$ -- i.e., the number of employed DTs -- and $k$ -- i.e., the number of neighbors considered to classify a sample.

In order to carry out the experimentation, we crafted two datasets, QB-CTU13 and EQB-CTU13. These two new sets were based on the CTU-13 dataset and designed to overcome its limitations regarding the imbalance of its classes, adding three new botnet classes, Bunitu, Miuref and NotPetya, to compare our most optimal model with other works.

Initially, we selected eleven of the most used features, we ranked the relevance of each one according to Gini Importance and Information Gain, and we obtained a list of the most informative features from the point of view of each method based on the information present in QB-CTU13 dataset.

Using this dataset, we analyzed the progression of the F1 score by adding features in the order established by each of the two rankings. We observed that seven features produced the biggest increase in the F1 score. However, we still evaluated three subsets of five, six and seven features to determine which one offered the best balance between computational time and F1 score. Regarding the F1 evolution, we noticed that changes in $m$ and $k$ parameters did not make a significant difference. Therefore we selected $m=10$ and $k=1$ because they were the least computationally expensive. 

After selecting the three feature subsets and the optimal parameters for the models, we tested them with the QB-CTU13 dataset. The evaluation was focused on the performance, stated as the ratio between the weighted F1 score and the computational time required to classify a sample. According to the results, DT with the 5-feature subset reached the best trade-off between detection ability and computational cost, surpassing the other two models and the other two feature subsets.

We compared the selected model, DT with the 5-feature subset, with another DT using all the eleven features. For this comparison, we employed the EQB-CTU13 dataset because it includes three more botnet classes, Bunitu, NotPetya and Miuref, allowing us to compare our models with other recent works. We found that the 5-feature subset achieved better F1 results than the whole feature set in three classes: Donbot, Murlo and NotPetya. Additionally, the recall of the 5-features DT on the Miuref and NotPetya classes -- i.e., $99\%$ and $97\%$ -- was significantly higher than the recall reported in the literature. The $90\%$ of F1 score for the Bunitu class reported in the found literature was not surpassed by the DT with the optimized feature subset, with only an F1 of $84\%$, but it was overcome by our model with eleven features using DT, with a $94\%$ of F1-score. 

Based on these results, we consider that our proposal using DT with the five selected features is the best solution for a multi-class botnet detector focused on the performance. It is true that using a single DT could produce over-fitting and, in some situations, it could be better to sacrifice time performance and use a Random Forest. However, in our proposal, we avoided over-fitting using a small number of features and testing the model with a cross-validation scheme.

We also compared our proposal with two SotA models: the SVM proposed by Joshi et al. \cite{joshi2020analysis} and the k-NN proposed by Ismail et al. \cite{ismail2021effects}. In contrast with our proposal, both models rely on the protocol and the IPs and as features. We did not consider the protocol since we already used the ports, which are usually used to identify the protocol, especially if the connection is ciphered with SSL/TLS or similar. We also did not consider the IP as features for three reasons: first, the IPs are usually assigned to the device by a DHCP service or are pre-configured, and thus the malware cannot change them and so the IP should not give information about the malware.  Second, the captured IPs might be masked by a NAT service or a proxy service. And third, we consider that since a given IP identifies a computer, then a model trained with this kind of feature could be over-fitted to the used dataset. For example, if in the dataset we have a computer with IP $10.0.0.5$ and it is infected with Miuref, then the model would learn that "any computer with IP $10.0.0.5$ is infected with Miuref". This statement is false, since in a different intranet a computer assigned with the IP $10.0.0.5$ can be clean of viruses. As future work, we plan to conduct an experiment to provide further evidence of this situation.

Finally, we want to remark that the obtained results show that k-NN is not suitable for this problem since its computational cost increases considerably with the number of samples used in the training phase. Even using just two features, the model was about $100$ times slower than DT using five features. 


Thinking in future works, we consider that a study on botnet detection over high bandwidth traffic could be interesting. In this scenario, the large number of packets sent by second makes it near impossible to analyze them all in real-time without dropping packets. Therefore, it will be necessary to determine the minimal requirements to detect botnet traffic in the hardest conditions. Additionally, we consider that the model performance can be further improved by implementing it using a compiled language such as C/C++ or Rust.

\footnotesize
\section*{Acknowledgements}

This work was supported by the framework agreement between the University of Le\'on and INCIBE (Spanish National Cybersecurity Institute) under Addendum 01,  and the FPU (Formaci\'on de Profesorado Universitario) grant of the Spanish Government with reference FPU18/05804.

\normalsize
\bibliography{references}

\end{document}